\documentstyle[aps,twocolumn,epsf]{revtex}
\title{Photodissociation in Quantum Chaotic Systems:
Random Matrix Theory of Cross-Section Fluctuations}
\author{
Yan V. Fyodorov$^1$\cite{leave}
and Y.  Alhassid$^2$       }
\address{
$^1$Fachbereich Physik, Universit\"at-GH Essen,
D-45117 Essen, Germany
        }
\address{$^2$ Center for Theoretical Physics, Sloane Physics
Laboratory, Yale University, New Haven, Connecticut 06520
        }
\date{\today}
\begin{document}
\draft
\maketitle
\bigskip

\begin{abstract}
Using  the random matrix description of open quantum chaotic systems we
calculate in closed form the universal autocorrelation function and the
probability distribution of the total photodissociation cross section in the
regime of quantum chaos.

\end{abstract}
\vfill

Recent advances in laser techniques allow  precision
measurements of the photodissociation of  polyatomic molecules \cite{Schrev}.
 Realistic quantum calculations of the
 photodissociation spectra for small polyatomic molecules,  such as the
 radicals HO$_2$ and NO$_2$ \cite{Schrev} or  the H$_3^{+}$ molecular ion
\cite{Mand} are also becoming available.
 The photodissociation describes the breakup of a bound molecule
 by the absorption of photons, and usually  proceeds
 through excited intermediate resonance states.
These resonance states are directly coupled to the continuum states that
describe possible breakup channels of the molecule into fragments,  and
their properties are expected to play a major role in the dissociation process.
Resonances are characteristic of all kinds of open systems, i.e.
systems in which motion along some directions is unbound.

One of the important observables is the total photodissociation cross-section.
 Often  it exhibits  irregular fluctuations consisting of  partly
overlapping peaks (e.g.  see \cite{Schrev} for the molecule
  HO$_2$) . Similar fluctuation patterns were first observed in
resonance neutron scattering \cite{Porter} and  are typical
 for many other systems such as  heavy ions\cite{ion} and atoms
\cite{Main}.    The statistics of the neutron resonance data  was
 explained by random matrix theory (RMT) \cite{Porter}, where the Hamiltonian
 is assumed to satisfy the underlying fundamental symmetries of the
 system but is otherwise random.  Originally justified
 by the complexity of the nuclear system, RMT  was later understood
to be applicable in systems whose underlying  classical dynamics is
 chaotic \cite{Bohigas}. Such chaotic systems are expected to
 exhibit  fluctuations whose statistical properties
are universal, i.e.  common to systems of different
physical nature and depending only on the symmetry class.
   RMT was successfully applied in the description of both bound and
 open quantum chaotic systems \cite{VWZ}.  Several universal features
of  chaotic scattering were recently calculated in closed form;
see Ref. \cite{FS} for  details.
We therefore expect that  the photodissociation cross-section of
 chaotic systems will  also display universal features.
Here we derive closed expressions for the
the autocorrelation function and  distribution of  the total
photodissociation cross-section in open chaotic systems.

 Following  the absorption of a photon, the excited molecule can dissociate
 into several channels.   A channel   describes  a  fragmentation
 of the system into several (possibly excited) fragments whose
 relative motion is described by a superposition of incoming and outgoing
 spherical waves.   We assume that at given energy $E$ there are  $M$
 different possible open channels.
 A dissociation solution $\Phi_c(E)$ to  the Schr\"{o}dinger equation
  is defined as a solution that satisfies the following boundary conditions:
an outgoing wave in exactly one open channel $c$, and incoming
 waves in all channels.    At the given energy $E$ there are exactly
 $M$ independent dissociation solutions $\Phi_c$  ($c=1,2,\ldots,M$).
The total  cross section $\sigma(E)$  for the molecule in its ground state
 $|g\rangle$ (or more generally any bound state)  to absorb a photon of
 energy $E-E_g$ and to dissociate
into any of its $M$ open channels is given, in the dipole approximation, by

\begin{equation}\label{def}
\sigma(E)=\sigma_0\sum_c\left|\left\langle  \Phi_c(E) |{\bf
\mu}|g \right\rangle\right|^2
\;,
\end{equation}
Here $\hat{\mu}=\mbox{\boldmath$\mu$}\cdot \hat{\bf e}$ is the 
component of dipole moment
$\mbox{\boldmath$\mu$}$ of the system along the polarization  $\hat{\bf e}$ of
 the absorbed light, and $\sigma_0= (2 \hbar^2 \epsilon_0 c)^{-1} (E -E_g)$.

To incorporate RMT description into scattering theory \cite{VWZ,FS}
it is convenient to divide the Hilbert space of the
dissociating system into two parts \cite{Wigner}: the internal
  ``interaction'' region, and the
external ``channel''   region  where the fragments are far enough from
each other that their interaction can be neglected.
Any  solution  $\Phi(E)$ at energy $E$ can then be
represented  in terms of its components
 ${\bf u}$ and $\psi$,
inside the interaction region  and channel region, respectively.
Using standard methods of  scattering theory  (see e.g. \cite{FS,Sok})
 one can relate the $M$ outgoing amplitudes of $\psi$ (denoted by the
 $M$-component vector ${\bf B}$) to the inside components ${\bf u}$
 by a linear  relation  ${\bf u}={\bf \hat{C}}{\bf B}$,
 where
\begin{equation}\label{int}
 {\bf \hat{C}}=\left(E- H_{in}-i\pi
WW^{\dagger}\right)^{-1} W \;.
\end{equation}
Here $H_{in}$ is the Hamiltonian describing the {\it closed} interaction
region when it is decoupled from the channel region
 (e.g. by imposing appropriate boundary
conditions), and the operator $W$ describes the coupling between the 
two parts of  the Hilbert space.

Assuming that the classical dynamics of the closed interaction region is
fully chaotic, we can replace the actual Hamiltonian
$H_{in}$ by a random $N\times N$ matrix taken from the Gaussian
 orthogonal ensemble (GOE) for systems with preserved time-reversal
invariance and from the Gaussian unitary ensemble (GUE) for systems
 with broken time-reversal  symmetry.
 The coupling $W$ is represented by an $N\times M$ matrix which we
 consider  to be fixed.  Because of the  invariance
 of the random matrix Hamiltonian $H_{in}$ under orthogonal (unitary)
transformations, the coupling  to the channels is essentially
 characterized by only $M$ invariants,
 the eigenvalues of $W^\dagger W$.  These eigenvalues can be
 expressed in term of the  transmission coefficients $T_c$
($0<T_c<1$),   defined through the
 averaged $S$-matrix by $T_c=1-|\langle S_{cc}\rangle|^2$. The limit
$T_c\ll 1$ corresponds to an almost closed channel $c$, whereas
 $T_c=1$ corresponds to the limit of  perfect coupling between the
interaction region and the channel $c$.

A central assumption in our model is that  direct transitions from the
ground state to the channels induced by the transition operator $\hat{\mu}$
 are negligible, and thus the decay is possible only  via  the
 excited resonance levels. Using this fact and
Eq. (\ref{int}),  the total photodissociation
 cross-section (\ref{def}) can be rewritten in the following optical theorem 
form:
\begin{equation}\label{im}
\sigma(E)\propto\left\langle g\left| \hat{\mu} {\bf \hat{C}}{\bf
\hat{C}^{\dagger}} \hat{
\mu} \right|g\right\rangle
\propto {\rm Im} \left\langle g\left| \hat{\mu}\frac{1}{E-{\cal H}_{eff}} \hat{
\mu} \right|g\right\rangle
\;,
\end{equation}
where ${\cal H}_{eff}=H_{in}-i\pi WW^{\dagger}$
is the effective non-Hermitean Hamiltonian
known to describe open chaotic systems.  For closed system ($W=0$),
 Eq. (\ref{im}) reduces to the strength function of
 $\hat{\mu}$, whose statistical properties were studied in 
Refs. \cite{AL92,Tan}.   The particular
resolvent form (\ref{im}) has the advantage of being suitable
for application of   Efetov's supermatrix
formalism\cite{Efrev,VWZ}.   The detailed presentation of such a
calculation can be
found in \cite{FS}, and here  we  only present the final result for the
 autocorrelation function of the photodissociation cross-section
defined by
\begin{equation}\label{corf}
S\left(\omega=\pi\Omega/\Delta\right)=
\frac{\langle\sigma(E-\Omega/2)\sigma(E+\Omega/2)\rangle}
{\langle \sigma(E)\rangle^2}-1 \;,
\end{equation}
with $\Delta$ being the mean level spacing for the Hamiltonian $H_{in}$.
$S(\omega)$ is found to be a sum of two terms
$S(\omega)=S_{1}(\omega)+S_{2}(\omega)$, which for the GOE case are
given by
\begin{equation}\label{main}\begin{array}{c}
S_{1,2}(\omega)=\\ \displaystyle{
\int_{-1}^1d\lambda\int_{1}^{\infty}d\lambda_1\int_{1}^{\infty}d\lambda_2
\frac{\cos{[\omega(\lambda_1\lambda_2-\lambda)]}(1-\lambda^2)}
{[\lambda_1^2+\lambda_2^2+\lambda^2-2\lambda_1\lambda_2\lambda-1]^2}}
\\\\
\times
\displaystyle{f_{1,2}(\lambda,\lambda_1,\lambda_2) \prod_{c=1}^{M}
\frac{(g_c+\lambda)}
{[(g_c+\lambda_1\lambda_2)^2-(\lambda_1^2-1)(\lambda_2^2-1)]^{1/2}}} \;,
\end{array}
\end{equation}
where
$$f_1(\lambda,\lambda_1,\lambda_2)=2\lambda_1^2\lambda_2^2-\lambda_1^2
-\lambda_2^2-\lambda^2+1;$$
$$f_2(\lambda,\lambda_1,\lambda_2)=(\lambda_1\lambda_2-\lambda)^2 \;.$$
The parameters $g_c$ are related to the transmission coefficients by
 $g_c=2/T_c-1$.
We note that  each of the contributions
$S_{1,2}(\omega)$  represents an interesting object by itself;
 $S_2(\omega)$ coincides with the autocorrelation function
of the Wigner time delays (studied in \cite{FS,td1,td11,Eck}),
whereas $S_1(\omega)$ is related to the Fourier transform of the
 ``norm leakage'' out of the interaction
region. The latter quantity was introduced recently by Savin and Sokolov as
a characteristic of the process of quantum relaxation in chaotic systems
\cite{Dima}.

For the GUE case one finds $S(\omega)$ to be
\begin{equation}\label{uni}
S(\omega)=\int_{-1}^1d\lambda\int_{1}^{\infty}d\lambda_1
\frac{\lambda_1}{\lambda_1-\lambda}\prod_c
\left(\frac{g_c+\lambda}{g_c+\lambda_1}\right)
\cos{[\omega(\lambda_1-\lambda)]} \;.
\end{equation}

In the limit  of a closed system  ($T_c=0$ for all $c$),  the  expressions
  (\ref{main}) and (\ref{uni}) reduce to the strength function correlators
calculated earlier  using RMT \cite{AL92} and  the supersymmetry
 method \cite{Tan}. This limit corresponds physically to
 energies $E$  below the threshold for
photodissociation (bound-to-bound transitions).
Simpler analytic forms can be obtained
in various limits, such as the limit of almost closed systems
 (i.e. all $T_c \ll 1$), and we defer the analysis to a future
publication. Here we discuss in detail only
 the limit of a large number of equivalent dissociation channels
 ($M\gg 1$, $T_c=T$) and large density of resonances $\rho=1/\Delta$,
 where  all resonances are found to have the same width
 $\Gamma$ \cite{FS}.  Such an
open system is characterized by a single parameter
$MT=2\pi\rho\Gamma\equiv\kappa$,
measuring the degree of resonance overlap  \cite{td1,FS,Dima}.
In this limit of homogeneously broadened resonances one finds
 (cf. \cite{td1}):
\begin{eqnarray}\label{S2w}
S_1(\omega)&=&\frac{2}{\beta}\frac{\kappa/2}{\omega^2+\kappa^2/4}\\
S_2(\omega)&=&\frac{\kappa/2}{\omega^2+\kappa^2/4}-\int_{-\infty}^{\infty}
d\tilde{\omega}Y_{2,\beta}(\tilde{\omega})\frac{\kappa/(2\pi)}{(\omega-
\tilde{\omega})^2+\kappa^2/4} \;,
\end{eqnarray}
where $Y_{2,\beta}(\omega)$ is Dyson's two-level correlation
function and $\beta=1,2$ for the GOE and GUE cases, respectively.
For strongly overlapping resonances, i.e. $\kappa\gg 1$,
the function $S_2(\omega)$  further simplifies
\begin{equation}\label{semi}
S_2(\omega)=\frac{1}{\beta}\frac{(\kappa/2)^2-\omega^2}
{\left[\omega^2+(\kappa/2)^2\right]^2} \;,
\end{equation}
and is only a $\kappa^{-1}$-order correction to $S_1(\omega)$.
Generically, for many open channels the
dominant  part of the autocorrelation function  for $\omega\sim\kappa$ is
always a Lorentzian.
For  $\kappa \ll 1$ the tail of $S(\omega)$ crosses the $\omega$ axis at
 $|\omega_0|\propto \kappa^{1/2}\gg\kappa$ and then approaches zero
from below.
 We note that  in  numerical simulations  of the chaotic photoionization of
hydrogen atom in external fields the cross-section autocorrelation was
 indeed found to be close to a
Lorentzian \cite{Main}.  In fact, for
$M\gg 1$ and  $\kappa\gg 1 $ it should be possible to
 apply semiclassical  considerations
 and  derive the autocorrelation function $S(\omega)$
 from the Gutzwiller trace formula,  as was done for
 the time-delay correlations \cite{Eck,FS}.

\begin{figure}
\epsfxsize= 7 cm
\centerline{\epsffile{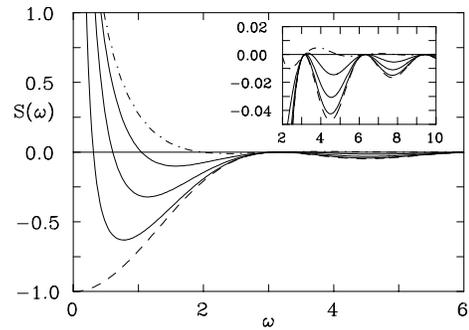}}
\caption
{ The autocorrelation function $S(\omega)$ of the photodissociation
 cross-section vs. $\omega$  for  $g \to \infty$ (closed system,
 dashed line), $g=20, 5, 2$ (solid lines) and $g =1$ (perfect transmission,
 dot-dashed line).  As the system opens  ($g$ decreases) the minimum
 gets shallower. The inset shows a magnification of the oscillatory behavior
 of  the decaying correlation.
}
\label{fig1}
\end{figure}

We emphasize, however, that  the autocorrelation
function for few open channels can differ substantially from a Lorentzian.
For that purpose, it is instructive to consider the single-channel case
 $M=1$ where Eq. (\ref{uni}) can be reduced to the  form:
\begin{equation}\label{M1}
S(\omega)=g\frac{\sin{2\omega}}{\omega}I_1(\omega)+
\frac{\sin^2{\omega}}{\omega^2}\left(1-2g\omega
I_2(\omega)\right)
\end{equation}
with
\begin{equation}
I_{1}(\omega)=\int_{0}^{\infty}dt\frac{\cos{\omega t}}{g+1+t};\quad
I_{2}(\omega)=\int_{0}^{\infty}dt\frac{\sin{\omega t}}{g+1+t} \;.
\end{equation}
This  $S(\omega)$  diverges logarithmically for small $\omega$, dips to
 a minimum below zero (``correlation hole'' \cite{AL92}) and then exhibits
 oscillatory decay to zero.
  As we open the system (i.e. $T$ increases from 0 to 1),  the correlation
hole gradually disappears and the amplitude of oscillations diminishes.
 Fig.  \ref{fig1} shows $S(\omega)$ for several values of $g$.
 In the limit $g \to \infty$, (\ref{M1}) reduces to $S(\omega) =
2\delta(\omega) - Y_2(\omega)$ \cite{AL92} (shown by the dashed line
without the $\delta$-function).

 Parametric correlations \cite{Tan} of the photodissociation
cross-sections can be
 calculated by incorporating the usual factor  of $\exp\left[-(x^2/2)
f_1(\lambda,\lambda_1,\lambda_2)\right]$
(GOE)  and $\exp \left[-x^2(\lambda_1^2-\lambda^2)\right]$ (GUE)
 into the integrand
of Eq.  (\ref{main}) and Eq. (\ref{uni}), respectively.

Another interesting quantity is the distribution function of the
scaled  photodissociation
cross-section  ${\cal P}(q)=\left\langle\delta\left(q- {\sigma(E)}/{\langle 
\sigma \rangle}\right)\right\rangle$.
To calculate ${\cal P}(q)$ we make use of the observation
 \cite{Missir,Sok} that  the $N\times N$  Hermitean
matrix ${\bf \hat{C}}{\bf \hat{C}^{\dagger}}$ has $N-M$ zero eigenvalues and
that its $M$ nonzero eigenvalues $\tau_c; \; c=1,...,M$ coincide with the
eigenvalues of $M\times M$ Wigner-Smith time delay matrix $Q={\bf
\hat{C}^{\dagger}}{\bf \hat{C}}$.
Denoting the eigenvectors of  ${\bf \hat{C}}{\bf \hat{C}^{\dagger}}$ that
correspond to $\tau_c$
by ${\bf u}_c$, we can  rewrite (\ref{im})  as
\begin{equation}\label{sig}
\sigma(E)\propto \sum_{c}\tau_c|V_c|^2;\quad V_c=\langle g|\hat{\mu}|{\bf
u}_c\rangle \;.
\end{equation}
This representation is useful in view of the known statistical properties
 of the time-delay matrix $Q$ \cite{FS,td1,td11,td2}.
In particular,  $V_c$, being proportional to the projection of ${\bf u}_c$
on the fixed vector $\hat{\mu} |g\rangle$, are  independent Gaussian variables
 with the same variance $\langle g |\hat{\mu}^2 |g\rangle/N$ ($V_c$
 are real for $\beta=1$ and complex for $\beta=2$) .
We note that since the  distribution of the normalized cross-section
 is independent of the fixed vector $\mu |g\rangle$,   ${\cal P}(q)$
coincides with that of the quantity $\rho_n(E)/ \langle\rho(E)\rangle$, where
$\rho_n(E)$ is local density of states
 (LDOS)  $\rho_n(E)= {\rm Im}  \langle
n|\left(E-{\cal H}_{eff}\right)^{-1}|n\rangle$ ($| n \rangle$ is a fixed
vector).  Special limiting cases of the LDOS distribution were
 studied in Refs. \cite{EP,Missir}.

In the particular case of one open channel the
 time-delay distribution ${\cal P}_\tau (\tau)$ is known
 in the whole crossover
regime between GOE and GUE symmetries\cite{td11}.
In the GOE and GUE limits ${\cal P}_\tau (\tau)$  simplifies to
 \cite{FS}:
\begin{eqnarray}\label{td}
{\cal P}_{\tau,\beta}(\tau)\propto\tau^{-\frac{\beta}{2}+2}\int_{0}^{\pi}d\phi
[g+\sqrt{g^2-1}\cos{\phi}]^{\beta/2}\\ \nonumber
\times \exp{-\frac{\beta}{2\tau}
[g+\sqrt{g^2-1}\cos{\phi}]} \;.
\end{eqnarray}
Using (\ref{td}) together with the known Gaussian  statistics of $V_c$
and the statistical independence of $\tau_c$ and $ {\bf u}_c$, we find from
(\ref{sig})
\begin{eqnarray}\label{m1}
{\cal P}_{\beta}(q)\propto\int_0^{\infty}
\frac{du}{u^{3-\beta}}e^{-\frac{\beta u^2}{2}}{\cal
P}_{\tau,\beta}\left(\frac{q}{u^2}\right)
\\
\propto \frac{1}{ q^{1-\beta/2}}\int_{0}^{\pi}d\phi
\frac{[g+\sqrt{g^2-1}\cos{\phi}]^{\beta/2}}
{\left(q+[g+\sqrt{g^2-1}\cos{\phi}]\right)^{\beta+1}} \;.
\end{eqnarray}
For $\beta=1$ this expression is equivalent to that obtained in
\cite{Missir} by a different method.

For the general case of $M>1$ open channels,  the distribution
of time-delays $\tau_c$ is known only for
the special case of perfect transmission (all $T_c=1$) \cite{td2}. In this case
 the LDOS distribution  turns out to be intriguingly
simple\cite{Missir}:
\begin{equation}\label{g1}
{\cal P}(q)\propto \frac{q^{\frac{\beta M}{2}-1}}{(1+q)^{\beta M+1}} \;.
\end{equation}

An alternative way to calculate the LDOS distribution is
 using the supersymmetry approach \cite{EP}, in which the general
 case  of arbitrary transmission coefficients  can be evaluated
for  systems with broken time-reversal symmetry:
\begin{eqnarray}\label{Pdist}
\nonumber {\cal P}(q)=\delta(q-1)+&&\\
\frac{1}{4\pi}\frac{\partial^2}{\partial q^2}\left[
(2q)^{1/2}\int_{q_{eff}}^{\infty}d\lambda_1\int_{-1}^{1}
\frac{d\lambda}{(\lambda_1-\lambda)} \right.&&\\ \nonumber
\frac{1}{(\lambda_1-q_{eff})^{1/2}}\left.
\prod_c\left(\frac{g_c+\lambda}{g_c+\lambda_1}\right)\right]&& \;,
\end{eqnarray}
where $q_{eff}=\frac{1}{2}\left(q+q^{-1}\right)$.
The integration  in (\ref{Pdist}) can be performed explicitly for an
arbitrary set of transmission coefficients.  For
equivalent channels ($g_c=g$) one finds a  rather simple result:
\begin{equation}\label{ge}
{\cal P}(q)=\frac{q^{M-1}}{(q^2+2gq+1)^{M+3/2}}\left[A_M(q^2+1)+q B_M\right] \;,
\end{equation}
where the coefficients $A_M,B_M$ are given by:
\begin{eqnarray}
A_M & = &\frac{1}{2}\left[(g+1)^{M+1}-(g-1)^{M+1}\right]
\frac{(2M-1)!!}{(M-1)!} \nonumber \\
B_M & = &\frac{1}{2}\left[(g+1)^{M+1}+(g-1)^{M+1}\right]
\frac{(2M+1)!!}{M!}-g
\frac{A_M}{M} \;.
\end{eqnarray}

For the case of perfect transmission $g=1$, we  reproduce Eq. (\ref{g1}).
In the opposite limit of  an almost closed
system  ($g\to\infty$),  we distinguish between two cases: i)  
 the regime of isolated resonances  ($g\to\infty$ at fixed $M$),  and
(ii) the regime of homogeneously broadened resonances  ($g\to\infty$
and $M\to\infty$ in such a way that $M/g=\kappa/2$ is constant).

In case (i) we distinguish  three different  regimes for $q$.
In the most important intermediate regime $1/g\ll q\ll g/M$, the distribution
 is independent
of the number of open channels $M$: ${\cal P}(q) \propto g^{-1/2}q^{-3/2}$,
whereas the distribution is $M$-dependent  for very large and very small
 values of $q$:
 ${\cal
P}(q) \propto g^{M}q^{-M+2}$ for  $q\gg g/M$, and
${\cal P}(q) \propto g^M q^{M-1}$ for $q\ll 1/g$.
In  case (ii) one can use the Stirling formula in Eq. (\ref{ge})
to derive  the distribution found in Refs. \cite{EP}.

 Except for the one-channel case (\ref{m1}),  we are not yet able to
 derive a general formula for the
 GOE case (i.e. an equation analogous to  (\ref{ge})).
However, we argue that in the limit (i) of  sharply non-overlapping
resonances
 ($g\gg M\sim 1$)  the correct formula can be obtained by replacing in our GUE
  formula $M\to M/2$. The  ``superuniversal'' law ${\cal P}(q) \propto
g^{-1/2}q^{-3/2}$ is similar to that found earlier for
the time-delay distribution \cite{FS,td11}, and  is expected to be
a generic feature of weakly open chaotic systems.

In conclusion,
we have calculated in closed form the universal autocorrelation
 function and the probability distribution of the total
photodissociation cross-section in the regime of quantum chaos.
Our  main assumptions are the applicability of
RMT for describing the closed
counterpart of the open quantum chaotic system, and the absence of  direct
field-induced transitions from the ground state to the continuum.
  In fact, our results hold for  the bound-to-continuum strength function
of an arbitrary transition operator and for any bound initial state, as
 long as the excited resonance states are fully chaotic.

Y.F.  acknowledges  V. Sokolov, D.Savin and J. Main for useful discussions.
This work was supported in part  by  SFB 237  ``Unordnung und grosse
Fluktuationen'' , and by the Department of Energy grant
No. DE-FG-0291-ER-40608.  Y.V.F. is grateful to
the Center for Theoretical Physics at Yale  University, and to the
 Newton Institute of
Mathematics at Cambridge, where part of this work was done,
 for the warm hospitality extended to him during his visits.

\begin{references}
\bibitem[*]{leave} On leave from Petersburg Nuclear Physics Institute,
Gatchina  188350, Russia.
\bibitem{Schrev} R. Schinke, H.-M. Keller, M. Stumpf and A.J. Dobbyn,
J.Phys.B: At.Mol. {\bf 28}, 2928 (1995).
\bibitem{Mand} V.A. Mandelshtam and H.S. Taylor,  J. Chem. Soc.,
Faraday Trans., {\bf 93}, 847 (1997).
\bibitem{Porter} C. E. Porter, {\it Statistical Theory of Spectra: Fluctuations}
(Academic Press, New York, 1965).
\bibitem{ion} M.Baldo, E.G.Lanza, A.Rapisarda, Chaos {\bf 3}, 691 (1993).
\bibitem{Main} J. Main and G. Wunner, J.Phys.B: At. Mol. {\bf 27}, 1994
(1994);
V.V. Flambaum, A.A. Gribakina and G.F. Gribakin,
  Phys. Rev. A  {\bf 54}, 2066 (1996)
\bibitem{Bohigas} O. Bohigas, in {\it Chaos and Quantum Physics},
 Les-Houches Session LII, ed. by M.J. Giannoni et.al (North
Holland, Amsterdam, 1991), p. 91.
\bibitem{VWZ}  J.J.M. Verbaarschot, H.A. Weidenm\"{u}ller and M.R.
Zirnbauer,  Phys. Rep. {\bf 129}, 367 (1985).
\bibitem{FS} Y.V. Fyodorov and H.-J. Sommers,  J. Math. Phys.  {\bf 38},1918
(1997),
and references therein.
\bibitem{Wigner} E.P. Wigner and L. Eisenbud, Phys. Rev. {\bf 72}, 29 (1947).
\bibitem{Sok} V.V. Sokolov and V.G. Zelevinsky,  Phys. Rev. C  {\bf 56},
311 (1997).
\bibitem{AL92} Y. Alhassid and R.D. Levine, Phys. Rev.  A {\bf  46}, 4650
 (1992); Y. Alhassid and N. Whelan,  Phys. Rev. Lett.  {\bf 70}, 572 (1993).
\bibitem{Tan} N. Taniguchi, A.V. Andreev and B.I. Altshuler,
Europh. Lett. {\bf 29}, 515 (1995);  N. Taniguchi and V.N.  Prigodin
Phys. Rev. B {\bf 54}, 14305  (1996).
\bibitem{Efrev} K.B. Efetov, Adv. Phys. {\bf 32}, 53 (1983).
\bibitem{td1}  N. Lehmann , D. Savin, V.V. Sokolov and H.-J. Sommers,
 Physica D {\bf 86},  572 (1995).
\bibitem{td11}Y.V.  Fyodorov, D. Savin and
H.-J. Sommers ,  Phys. Rev. E {\bf 55}, 4857 (1997).
\bibitem{Eck}B. Eckhardt, Chaos {\bf 3}, 613 (1993).
\bibitem{Dima}  D.V. Savin and V.V. Sokolov,  Phys. Rev. E
{\bf 56}, R4911 (1997).
\bibitem{EP} K.B. Efetov and V.N. Prigodin,  Phys. Rev. Lett.  {\bf 70},
1315 (1993);  C.W. Beenakker, Phys. Rev. B {\bf 50}, 15170 (1994);
A.D. Mirlin and Y.V. Fyodorov, Europh. Lett. {\bf 25}, 669 (1994).
\bibitem{Missir} T.Sh. Misirpashaev, P.W. Brouwer and
 C.W.J. Beenakker, Phys. Rev. Lett. {\bf 79}, 1841 (1997).
\bibitem{td2} P.W. Brouwer, K.M. Frahm and  C.W.J. Beenakker,
Phys. Rev. Lett. {\bf 78}, 4737 (1997);
V. Gopar, P.A. Mello and M. B\"{u}ttiker, Phys. Rev. Lett. 
{\bf 77}, 3005 (1996).
\end{references}
\end{document}